\documentstyle[11pt,aaspp4]{article}
\def\deg{$^{\rm o}$}

\def\kms{\ifmmode {\rm km\ s}^{-1} \else km s$^{-1}$\fi}
\def\Msun{\ifmmode M_{\odot} \else $M_{\odot}$\fi}
\def\Lsun{\ifmmode L_{\odot} \else $L_{\odot}$\fi}
\def\Rs{\ifmmode R_{\rm S} \else $R_{\rm S}$\fi}
\def\qo{\ifmmode q_{\rm o} \else $q_{\rm o}$\fi}
\def\Ho{\ifmmode H_{\rm o} \else $H_{\rm o}$\fi}
\def\ho{\ifmmode h_{\rm o} \else $h_{\rm o}$\fi}
\def\ltsim{\raisebox{-.5ex}{$\;\stackrel{<}{\sim}\;$}}

\def\vFWHM{\ifmmode V_{\mbox{\tiny FWHM}} \else
            $V_{\mbox{\tiny FWHM}}$\fi}
\def\CCF{\ifmmode F_{\it CCF} \else $F_{\it CCF}$\fi}
\def\ACF{\ifmmode F_{\it ACF} \else $F_{\it ACF}$\fi}
\def\Halpha{\ifmmode {\rm H}\alpha \else H$\alpha$\fi}
\def\Hbeta{\ifmmode {\rm H}\beta \else H$\beta$\fi}
\def\Hgamma{\ifmmode {\rm H}\gamma \else H$\gamma$\fi}
\def\Hdelta{\ifmmode {\rm H}\delta \else H$\delta$\fi}
\def\Lya{\ifmmode {\rm Ly}\alpha \else Ly$\alpha$\fi}
\def\Lyb{\ifmmode {\rm Ly}\beta \else Ly$\beta$\fi}

\def\heii{He\,{\sc ii}}

\def\ciii{\ifmmode {\rm C}\,{\sc iii} \else C\,{\sc iii}\fi}
\def\civ{\ifmmode {\rm C}\,{\sc iv} \else C\,{\sc iv}\fi}

\def\nv{N\,{\sc v}}
\def\oi{O\,{\sc i}}

\def\o5007{[O\,{\sc iii}]\,$\lambda5007$}

\def\mgii{Mg\,{\sc ii}}

\lefthead{Peterson and Wandel}
\righthead{Black Holes in AGNs}

\begin{document}
\title{Keplerian Motion of Broad-Line Region
Gas as Evidence for Supermassive Black Holes in Active Galactic Nuclei}

\author{
Bradley M. Peterson\altaffilmark{1}
and Amri Wandel\altaffilmark{2,3}
}
\altaffiltext{1}
           {Department of Astronomy, The Ohio State University,
    	174 West 18th Avenue, Columbus, OH 43210-1106\\
	Email: peterson@astronomy.ohio-state.edu}
\altaffiltext{2}
        {Department of Physics and Astronomy,
	University of California at Los Angeles,
	Los Angeles, CA 90095-1562\\
	Email: wandel@astro.ucla.edu}
\altaffiltext{3}
	{Permanent address:
	Racah Institute, The Hebrew University, Jerusalem 91904, ISRAEL}

\begin{abstract}
Emission-line variability data on NGC 5548 argue strongly for the
existence of a mass of order $7 \times 10^7$\,\Msun\
within the inner few light days of the nucleus in the
Seyfert 1 galaxy NGC 5548. The time-delayed response of the
emission lines to continuum variations is used to infer the
size of the line-emitting region, and these determinations are
combined with measurements of the Doppler widths of the
variable line components to estimate a virial mass. The
data for several different emission lines spanning an order of magnitude
in distance from the central source show the expected
$V \propto r^{-1/2}$ correlation and are consistent with a single value for
the mass. 
\end{abstract}

\keywords{galaxies: active --- galaxies: individual (NGC 5548) ---
galaxies: quasars: emission lines --- galaxies: Seyfert}
 
\setcounter{footnote}{0}

\section{Introduction}

Since the earliest days of quasar research, supermassive
black holes (SBHs) have been considered to be a likely, if not the
most likely, principal agent of the activity
in these sources. 
Evidence for the existence of SBHs
in active galactic nuclei (AGNs), and indeed in non-active nuclei as well, has
continued to accumulate (e.g., Kormendy \& Richstone 1995).
In the specific case of AGNs, probably the strongest
evidence to date for SBHs has been Keplerian motions
of megamaser sources in the Seyfert galaxy NGC~4258
(Miyoshi et al.\ 1995) and asymmetric Fe\,K$\alpha$ emission
in the X-ray spectra of AGNs (e.g., Tanaka et al.\ 1995), 
though the latter is still
somewhat controversial as the origin of the Fe\,K$\alpha$ emission
has not been settled definitively.

The kinematics of 
the broad-line region (BLR) potentially provide a means of measuring
the central masses of AGNs. A virial estimate of the central mass,
$M \approx r \sigma^2/G$, can be made by using the line 
velocity width $\sigma$, which is typically several thousands of
kilometers per second, and the size of the emission-line region
$r$. For this to be  meaningful, we must know that
the BLR gas motions are dominated by gravity,
and we must have some reliable estimate of the BLR size.
The size of the BLR can be measured by 
reverberation mapping (Blandford \& McKee 1982),
and this has been done for more than two dozen AGNs.
Whether or not the broad emission-line widths 
actually reflect virial motion is still somewhat problematic:
while the relative response time scales for the blueshifted and
redshifted wings of the lines reveal no strong signature of
outflow, there are still viable models
with non-gravatitionally driven cloud motions. However, if
the kinematics of the BLR can be proven to be gravitationally
dominated, then the BLR provides an even more definitive demonstration
of the existence of SBHs than megamaser kinematics because
the BLR is more than two orders of magnitude closer to the
central source than the megamaser sources.
Recent investigations of AGN virial masses estimates based
on BLR sizes have been quite promising 
(e.g., Wandel 1997; Laor 1998) and suggest that this method
ought to be pursued.

In this Letter, we argue that the broad emission-line
variability data on one of the best-studied AGNs,
the Seyfert 1 galaxy NGC 5548, demonstrates that
the BLR kinematics are Keplerian, i.e., that the
emission-line cloud velocities are dominated by
a central mass of order $7 \times 10^{7}$\,\Msun\
within the inner few light days 
($r \ltsim 5 \times 10^{15}$\,cm). We believe that
this strongly supports the hypothesis that
SBHs reside in the nuclei of active galaxies.

\section{Methodology}

Measurement of virial masses from
emission lines requires (1) determination of the BLR size,
(2) measurement of the emission-line velocity dispersion, 
and (3) a demonstration that the kinematics are gravitationally dominated.
A correlation between the BLR size and line-width of the form
$r \propto \sigma^{-2}$ is consistent with a wide variety of
gravitationally dominated kinematics. It thus provides good evidence
for such a dynamical scenario, although alternative pictures which
contrive to produce a similar result cannot be ruled out.
Indeed, the absence of such a relationship has been regarded as
the missing item in AGN SBH measurements (Richstone et al.\ 1998).

For gravitationally dominated dynamics,
the size--line-width relationship must hold for all lines at all times.
To test this, we consider the case of NGC 5548, which has been the subject
of extensive UV and optical monitoring campaigns by the
International AGN Watch consortium\footnote{Information about the
International AGN Watch and copies of published data
can be obtained on the World-Wide Web
at URL
{\sf http://www.astronomy.ohio-state.edu/$\sim$agnwatch/}.}
(Alloin et al.\ 1994)
for more than ten years. The data 
are from UV monitoring programs undertaken 
with the {\it International Ultraviolet Explorer (IUE)} 
in 1989 (Clavel et al.\ 1991) and 
with {\it IUE}\ and {\it Hubble Space Telescope (HST)} in
1993 (Korista et al.\ 1995),
plus ground-based spectroscopy from 1989 to 1996
(Peterson et al.\ 1999 and references therein). We
consider the response of a variety of lines in two separate
observing seasons (1989 and 1993) and the response of
\Hbeta\ over an eight-year period.

Cross-correlation of the continuum and emission-line light curves yields
a time delay or ``lag'' that is interpreted as the light-travel time
across the BLR. Specifically, 
the centroid of the cross-correlation function (CCF)
$\tau_{\rm cent}$ times the signal propagation speed $c$
is the  responsivity-weighted mean radius of the BLR 
for that particular emission line (Koratkar \& Gaskell 1991).

We have measured $\tau_{\rm cent}$ for various emission lines 
using light curves of NGC 5548 in the
AGN Watch data base and the interpolation cross-correlation  
method as described by White \& Peterson (1994).
The UV measurements for 1989 are the GEX values from 
Clavel et al.\ (1991). The UV measurements for 
1993 are taken from Tables 12--14 and 16--17 of Korista et al.\ (1995). The
optical data for 1989--1993 are from Wanders \& Peterson (1996)
and from Peterson et al.\ (1999) for 1994--1996.  
Uncertainties in these values were determined as described by
Peterson et al.\ (1998b).
The results are given in Table 1, in which columns (1) and (2)
give the epoch of the observations and
the emission line, respectively. Column (3) gives the lag $\tau_{\rm cent}$
and its associated uncertainties.

Emission-line widths are not simple to measure on account of
contamination by emission from the narrow-line region, and in
some cases, contamination from other broad lines.
We have circumvented this problem by using a large number
of individual spectra to compute mean and root-mean-square (rms)
spectra, and we measure the width of the emission features in
the rms spectrum. The advantage of this approach is that 
constant or slowly varying components of the spectrum do not
appear in the rms spectrum, and the emission features in
the rms spectrum accurately represent the parts of the emission
line that are varying, and for which the time delays
are measured (Peterson et al.\ 1998a). This
technique requires very homogeneous spectra:
for the 1989 UV spectrum, we used the GEX-extracted
SWP spectra. For the 1993 UV spectrum, we used the
{\it HST} FOS spectra, excluding those labeled ``dropouts''
by Korista et al.\ (1995) which were not optimally centered
in the FOS aperture.
For the optical spectra through 1993,
we used the homogeneous subset analyzed by Wanders \& Peterson (1996),
and a similar subset for 1994--1996.
In each rms spectrum, we determined
the full-width at half-maximum (FWHM) of each measurable line,
with a range of uncertainty estimated by the highest and lowest
plausible settings of the underlying continuum. 
The line widths are given as line-of-sight
Doppler widths in kilometers per second in column (4) of Table 1.

Each emission line provides an independent measurement of
the virial mass of the AGN in NGC 5548 by combining the
emission-line lag with its Doppler width in the rms spectrum.
Column (5) of Table 1 gives a virial mass estimate 
$M = f r_{\rm BLR} \sigma_{\rm rms}^2/G$ for
each line, where
$\sigma_{\rm rms} = \sqrt{3}\vFWHM/2$
(Netzer 1990) and $r_{\rm BLR} = c \tau_{\rm cent}$.
The factor $f$ depends on the details of the geometry, kinematics, and
orientation of the BLR, as well as the emission-line responsivity
of the individual clouds, and is expected to be of order
unity. Uncertainty in this factor limits the accuracy of
our mass estimate to about an order of magnitude
(see \S{3}). 
Neglecting the systematic uncertainty in $f$,
the unweighted mean of all these mass estimates
is $6.8\,(\pm 2.1) \times 10^{7}$\,\Msun. 
To within the quoted uncertainties, all of the mass measurements
are consistent. The large systematic uncertainty should not obscure 
the key result, namely that the quantity 
$r_{\rm BLR} \sigma_{\rm rms}^2/G$ is constant and
argues strongly for a central mass of order $7 \times 10^{7}$\,\Msun.

In Fig.\ 1, we show the measured emission-line lag $\tau_{\rm cent}$, 
plotted as a function of the width of the line in the
rms spectrum for various broad emission lines in NGC 5548.
Within the measurement uncertainties,
all the lines yield identical values for
the central mass.  A weighted fit to the relationship 
$\log (\tau_{\rm cent}) = a + b\log (\vFWHM)$ yields
$b=-1.96\pm0.18$, consistent with the expected value
$b=-2$, although the somewhat high reduced $\chi^2_{\nu}$ value of 1.70
(compared with $\chi^2_{\nu} = 2.14$ for a
forced $b = -2$ fit as shown in the figure)
suggests that there may be additional sources of scatter in
this relationship beyond random error.

If our virial hypothesis is indeed correct, we should
measure the same mass using independent data obtained at different times.
The \Hbeta\ emission line in NGC 5548 is the only line
for which reverberation measurements have been made for
multiple epochs. In Fig.\ 2a,
we show the measured \Hbeta\ lag as a function of the width
of the \Hbeta\ line in the rms spectrum for the six years listed in
Table 1. 
The relationship is shallower than that seen in the multiple-line
data shown in Fig.\ 1 ($b=-0.72\pm0.29$ with $\chi^2_{\nu} = 0.79$),
and indeed is poorly fit with the expected virial slope
(for the $b=-2$ fit shown in the figure, $\chi^2_{\nu} = 3.71$, although 
more than 50\% of the contribution to $\chi^2_{\nu}$ 
is due to the single data point from 1996).
Note that data from two years,
1993 and 1995, have not been included in this plot because the rms
spectra for these two years have a strong double-peaked structure
that we are unable to account for at present. We also note that
a rather better relationship between the \Hbeta\ time lag
and rms line width is found if we use the CCF peak rather than
the centroid for the BLR size, as shown in Fig.\ 2b
($b=-1.47\pm0.21$ with $\chi^2_{\nu} = 0.59$, and for
the $b=-2$ fit shown in the figure, $\chi^2_{\nu} = 1.58$). The
CCF centroid represents the responsivity-weighted mean radius
of the \Hbeta\ line-emitting region, but the CCF peak has no
similarly obvious interpretation, though in some geometries the 
cross-correlation peak is a probe of the emission-line gas
closest to the central source.
In any case, the virial mass we infer from the mean of the \Hbeta\ data 
is the same within the uncertainties
regardless of whether the CCF centroid 
($7.3\,(\pm2.0)\times 10^{7}$\,\Msun) or
peak ($6.8\,(\pm1.0)\times 10^{7}$\,\Msun)
is used to infer the BLR size. 
There are a number of possible reasons for the large
$\chi^2$ values for the virial fits; it is important to
remember that both the lag and line width are dynamic
quantities that are dependent on the mean continuum flux, which
can change significantly over the course of an observing season.
We attempted to test this by isolating individual ``events'' 
in the light curves and repeating the analysis. Unfortunately,
the relatively few spectra in each event significantly degraded
the quality of both the lag and line-width measurements and thus
proved to be unenlightening.

A diagram similar to our Fig.\ 1 was published by
Krolik et al.\ (1991) for NGC 5548. 
We believe that our improved treatment, plus additional
data, makes the case more compelling primarily because we
measured  the broad-line widths
from the variable part of the spectrum only (i.e., the rms spectrum)
rather than by multiple-component fitting of the broad-line profiles.
Also, we included only lines for which we could determine both accurate
lags and line widths in the rms spectra, thus excluding
\Lya\,$\lambda1215$ because of contamination
by geocoronal \Lya\ in the rms spectrum,
\nv\,$\lambda1240$ because it is weak and badly blended
with \Lya, and
\oi\,$\lambda1304$ on account of its low contrast in the rms
spectrum. We excluded \mgii\,$\lambda2798$ because of its
poorly defined time lag --- the response of this line is long
enough for aliasing to be a problem. Finally, we also
included optical lines (\Hbeta\ and \heii\,$\lambda4686$)
not included by Krolik et al., plus additional UV measurements
from the 1993 monitoring campaign.

An obvious question to ask is whether or not it is possible to
{\em directly} determine the BLR kinematics by differential
time delays between various parts of emission lines (e.g.,
in the case of purely radial infall, the redshifted side of an
emission line should respond to continuum changes before the
blueshifted side). In general, cross-correlations of
emission-line fluxes in restricted Doppler-velocity ranges
have failed to yield significant time lags in the several
AGNs tested to date (e.g., Korista et al.\ 1995), consistent
with, although not proving, the virial hypothesis.

\section{Discussion}
We have shown that the emission-line time-lag/velocity-width
relationship argues very strongly for
an SBH of mass $\sim7\times10^7\,$\Msun\ in the nucleus of NGC 5548. 
The accuracy of this determination is limited by unknown systematics
involving the geometry, kinematics, and line reprocessing physics of
the BLR. As a simple illustration, we consider \civ\,$\lambda1549$
line emission from a BLR consisting of clouds in a Keplerian disk
with radial responsivity proportional to $r^{-2.5}$ (which is steep
enough to make the results fairly insensitive to the outer radius
of the disk) and inner radius $R_{\rm in} = 3$\,lt--days.
A relatively low central mass ($5\times10^6$\,\Msun) 
with high inclination ($i=90$\deg) disk and asymmetric line emission
can fit the 1989 \civ\ results in Table 1.
At the other extreme, a larger mass ($1.1\times10^8$\,\Msun)
is required for a lower inclination ($i=20$\deg) and 
isotropic line emission. For further comparison, the
specific model of Wanders et al.\ (1995), based on anisotropically 
illuminated clouds in randomly inclined Keplerian orbits, requires
$M = 3.8 \times 10^{7}$\,\Msun, and extrapolation to the BLR
of the Fe K$\alpha$ disk model of Nandra et al.\ (1997) requires
$M = 3.4\times10^{7}$\,\Msun.

As shown by Peterson et al.\ (1999), the \Hbeta\ emission-line lag
varies with continuum flux, though as with the results discussed
here, the correlation shows considerable scatter, probably because
of the dynamic nature of the quantities being measured.
But it seems clear that as the continuum luminosity
increases, greater response is seen from gas at larger
distances from the central source. We argue here that this also results
in a change in the emission-line width; as the response
becomes dominated by gas further away from the central
source, the Doppler width of the responding
line becomes narrower. This shows
that the different widths of various emission lines
is related to the radial distribution of the line-emitting gas ---
high-ionization lines arise at small distances and have large
widths, and low-ionization lines arise primarily at larger
radii and are thus narrower.

While this accounts for some important
characteristics of AGN emission lines and their variability, it
is nevertheless clear that this is not the entire story;
there is still scatter in the relationships that is
unaccounted for by these correlations, and their are other
phenomena that are not accounted for in this simple interpretation.
For example, for central masses as large as reported here,
observable relative shifts in the positions of the emission
lines are expected from differential gravitational
redshifts. The gravitational redshift for each line in NGC 5548 is
given by
\begin{equation}
\Delta V =\frac{GM}{cr_{\rm BLR}} \approx 
\frac{1160\,\mbox{\kms}}{r_{\rm BLR}\,\mbox{\rm (light days)}}.
\end{equation}
This clearly predicts that that high-ionization lines ought to
be redshifted relative to the low-ionization lines, when in
fact the opposite is observed in higher-luminosity objects
(Gaskell 1982; Wilkes 1984).
However, the gravitational redshift in NGC 5548 should apply to the
{\it variable} component of the emission line only and would be
sufficiently small to be unobservable in our rms spectra.
The occasional appearance of double-peaked rms profiles
is yet another complication. As noted earlier, in two of the eight years
of optical data on NGC 5548, the \Hbeta\ profile in the rms spectrum
is strongly double-peaked. We do not see an obvious explanation
for why the emission line should be single-peaked on some occasions
and double-peaked on others.

\section{Summary}

We have shown that in the case of the Seyfert 1 galaxy NGC 5548
emission-line variability data yield a consistent virial mass
estimate $M \approx 7 \times 10^7$\,\Msun, though systematic
uncertainties about the BLR geometry, kinematics, and line-reprocessing
physics limit the accuracy of the mass determination to about
an order of magnitude. Data on
multiple emission lines spanning a factor of ten or more
in distance from the central source shows the 
$r_{\rm BLR} \propto \vFWHM^{-2}$ correlation expected
for virialized BLR motions. The time delay of \Hbeta\
emission is known to vary by at least a factor of two
over a decade (Peterson et al.\ 1999), and we show here
that the line-width variations
are anticorrelated with the time-delay variations.
The central mass is concentrated inside a few light days,
which corresponds to about 250 Schwarzschild radii ($R_{\rm S} = 2 GM/c^2$)
for the mass we infer, which argues very strongly for
the existence of an SBH in NGC 5548.

\acknowledgements{BMP is grateful for support of this work
by the National Science Foundation and NASA through grants
AST--9420080 and NAG5--8397, respectively,
to The Ohio State University. AW wishes to
acknowledge the hospitality of the Department of Physics and
Astronomy at UCLA during this work. We thank K.T.\ Korista,
M.A.\ Malkan, P.S.\ Osmer, R.W.\ Pogge, J.C.\ Shields,
and B.J.\ Wilkes for critical reading of the
draft manuscript and A.\ Gould for helpful advice.}


\clearpage

\begin{figure}
\caption{The time lags (cross-correlation function
centroids $\tau_{\rm cent}$) in days (1 light day = $2.6 \times 10^{15}$\,cm)
for various lines in NGC 5548 are plotted as a function
of the full-width at half maximum of the feature (in the rest
frame of NGC 5548) in the root-mean-square spectrum. Points
plotted are listed in Table 1. The filled circles refer to data from
1989, and the open circles to data from 1993.
The dotted line indicates a fixed virial mass 
$M = 6.8 \times 10^7$\,\Msun.}
\end{figure}

\begin{figure}
\caption{This plot shows six measurements (based on yearly averages)
of the broad \Hbeta\ emission-line time lag and corresponding width
of the \Hbeta\ feature in the rms spectrum formed from a
homogeneous subset of spectra obtained during each year.
In panel {\it a} (left), the time delay is based on the
centroid of the continuum--\Hbeta\ CCF, and in panel {\it b} (right),
the location of the peak value of the CCF is used. 
The dotted line corresponds to  a fixed virial mass 
$M = 6.8 \times 10^7$\,\Msun.}
\end{figure}

\clearpage


\begin{deluxetable}{llccc}
\tablewidth{0pt}
\tablecaption{Virial Mass Estimates for NGC 5548}
\label{tab:sourcetab}
\tablehead{
\colhead{ } &
\colhead{Emission} &
\colhead{$\tau_{\rm cent}$} &
\colhead{$V_{\mbox{\scriptsize FWHM}}$} &
\colhead{Mass} \nl
\colhead{Year} & 
\colhead{Line} &
\colhead{(days)} &
\colhead{(km s$^{-1}$)} &
\colhead{($10^7$\,\Msun)} \nl
\colhead{(1)} & 
\colhead{(2)} & 
\colhead{(3)} & 
\colhead{(4)} & 
\colhead{(5)} 
} 
\startdata
1989 & Si\,{\sc iv}$\lambda1400$ & 
	$12.0^{+4.4}_{-2.6}$ & $6320 \pm 1470$ & $7.0^{+4.2}_{-3.6}$ \nl
     & + O\,{\sc iv}]$\lambda1402$ & \nl
     & C\,{\sc iv}$\lambda1549$ & 
	$ 9.5^{+2.6}_{-1.0}$ & $5520 \pm 380$ & $4.2^{+1.3}_{-0.7}$ \nl
     & He\,{\sc ii}$\lambda1640$ & 
	$ 3.0^{+2.9}_{-1.1}$ & $8810 \pm 1800$ & $3.4^{+3.6}_{-1.9}$ \nl
     & C\,{\sc iii}]$\lambda1909$ & 
	$27.9^{+6.0}_{-5.5}$ & $4330 \pm 770$ & $7.7^{+3.2}_{-3.1}$ \nl
     & He\,{\sc ii}$\lambda4686$ & 
	$8.5^{+3.4}_{-3.4}$ & $8880 \pm 1510$ & $9.8^{+5.2}_{-5.2}$ \nl
     & H$\beta\,\lambda4861$ & 
	$19.7^{+2.0}_{-1.4}$ & $4250 \pm 240$ & $5.2^{+0.8}_{-0.7}$ \nl
1990 & H$\beta\,\lambda4861$ & 
	$19.3^{+1.9}_{-3.0}$ & $4850 \pm 300$ & $6.6^{+1.0}_{-1.3}$ \nl
1991 & H$\beta\,\lambda4861$ & 
	$16.4^{+3.8}_{-3.3}$ & $5700 \pm 480$ & $7.8^{+2.2}_{-2.0}$ \nl
1992 & H$\beta\,\lambda4861$ & 
	$11.4^{+2.3}_{-2.3}$ & $5830 \pm 300$ & $5.7^{+1.3}_{-1.3}$ \nl
1993 & Si\,{\sc iv}$\lambda1400$ & 
	$4.6^{+0.8}_{-1.4}$ & $9060 \pm 2320$ & $5.5^{+3.0}_{-3.3}$ \nl
     & + O\,{\sc iv}]$\lambda1402$ & \nl
     & C\,{\sc iv}$\lambda1549$ & 
	$ 6.8^{+1.1}_{-1.1}$ & $8950 \pm 570$ & $8.0^{+1.6}_{-1.6}$ \nl
     & He\,{\sc ii}$\lambda1640$ & 
	$ 2.0^{+0.3}_{-0.4}$ & $13130 \pm 4500$ & $5.1^{+3.5}_{-3.6}$ \nl
1994 & H$\beta\,\lambda4861$ & 
	$15.5^{+2.3}_{-6.1}$ & $6860 \pm 420$ & $10.7^{+2.1}_{-4.4}$ \nl
1996 & H$\beta\,\lambda4861$ & 
	$16.8^{+1.4}_{-1.4}$ & $5700 \pm 420$ & $8.0^{+1.4}_{-1.4}$ \nl
\enddata
\end{deluxetable}



\clearpage

\begin{references}
\reference{}Alloin, D., Clavel, J., Peterson, B.M., Reichert, G.A., \& 
Stirpe, G.M. 1994, in Frontiers of Space and Ground-Based Astronomy, ed.\ 
W. Wamsteker, M.S. Longair, \& Y. Kondo  (Dordrecht: Kluwer), p.\ 423
\reference{}Blandford, R.D., \& McKee, C.F. 1982, ApJ, 255, 419
\reference{}Clavel, J., et al. 1991, ApJ, 366, 64
\reference{}Gaskell, C.M. 1982, ApJ, 263, 79
\reference{}Koratkar, A.P., \& Gaskell, C.M. 1991, ApJS, 75, 719
\reference{}Kormendy, J., \& Richstone, D. 1995, ARAA, 33, 581
\reference{}Korista, K.T., et al. 1995, ApJS, 97, 285
\reference{}Krolik, J.H., Horne, K., Kallman, T.R., Malkan, M.A., 
Edelson, R.A., \& Kriss, G.A. 1991, ApJ, 371, 541
\reference{}Laor, A. 1998, ApJ, 505, L83
\reference{}Miyoshi, M., Moran, J., Herrnstein, J., Greenhill, L.,
Nakai, N., Diamond, P., \& Inoue, E. 1995, Nature, 373, 127
\reference{}Nandra, K., George, I.M., Mushotzky, R.F., 
Turner, T.J., Yaqoob, T. 1997, ApJ, 477, 602
\reference{}Netzer, H. 1990, in Active Galactic Nuclei, 
R.D.\ Blandford, H.\ Netzer, \& L.\ Woltjer (Berlin: Springer-Verlag),
p.\ 137
\reference{}Peterson, B.M., et al. 1999, ApJ,  510, 659
\reference{}Peterson, B.M., Wanders, I.,  Bertram, R.,
Hunley, J.F., Pogge, R.W.,  \& Wagner, R.M. 1998a, ApJ, 501, 82
\reference{}Peterson, B.M., Wanders, I., Horne, K., Collier, S.,
Alexander, T., \& Kaspi, S. 1998b, PASP, 110, 660
\reference{}Richstone, D., et al. 1998, Nature, 395, A14.
\reference{}Tanaka, Y., et al. 1995, Nature, 375, 659
\reference{}Wandel, A. 1997, ApJ, 490, L131
\reference{}Wanders, I., et al. 1995, ApJ, 453, L87
\reference{}Wanders, I., \& Peterson, B.M. 1996, ApJ, 466, 174
(Erratum: 1997, ApJ, 477, 990) 
\reference{}White, R.J., \& Peterson, B.M. 1994, PASP, 106, 879
\reference{}Wilkes, B.J. 1984, MNRAS, 207, 73
\end{references}
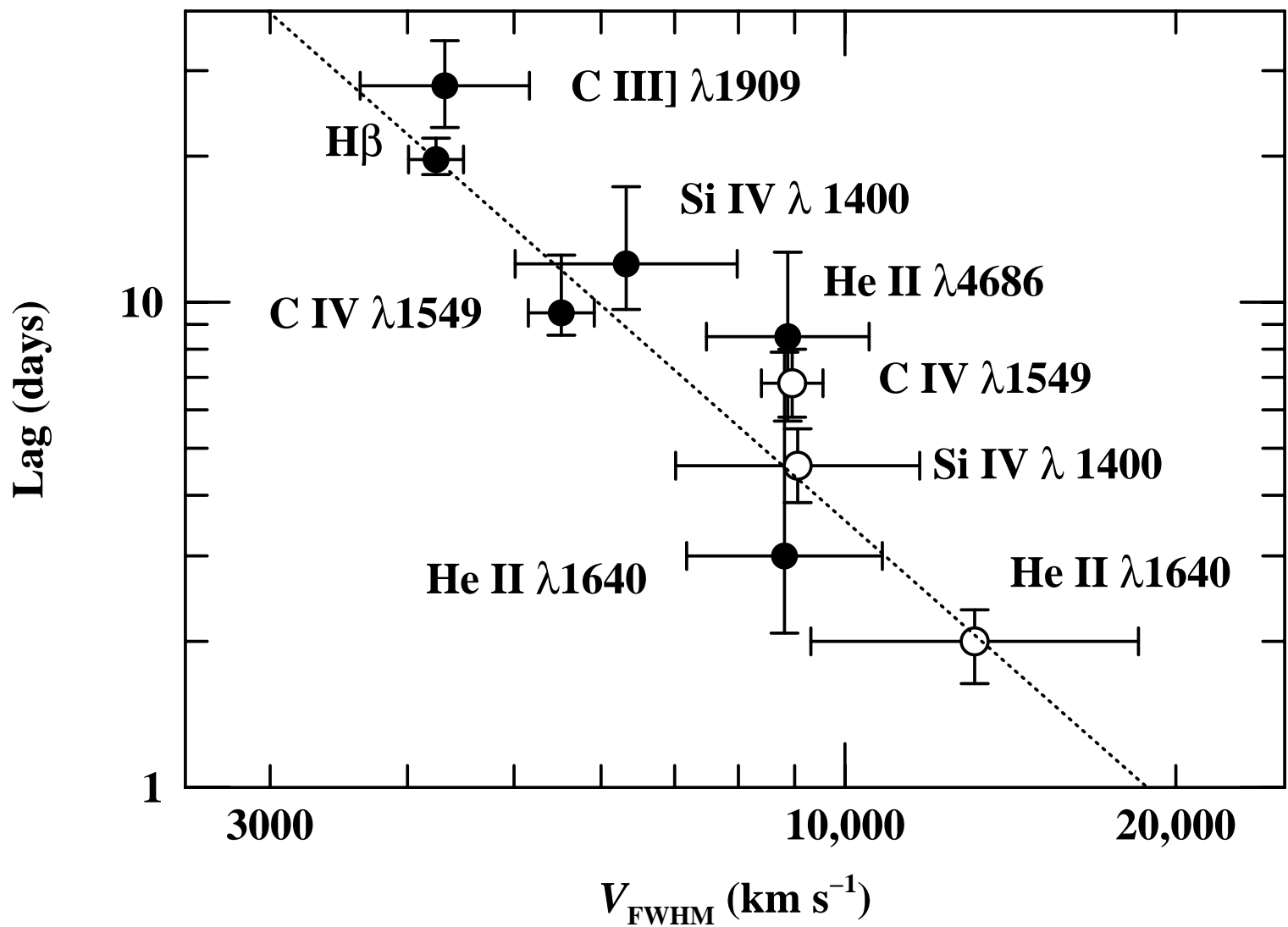
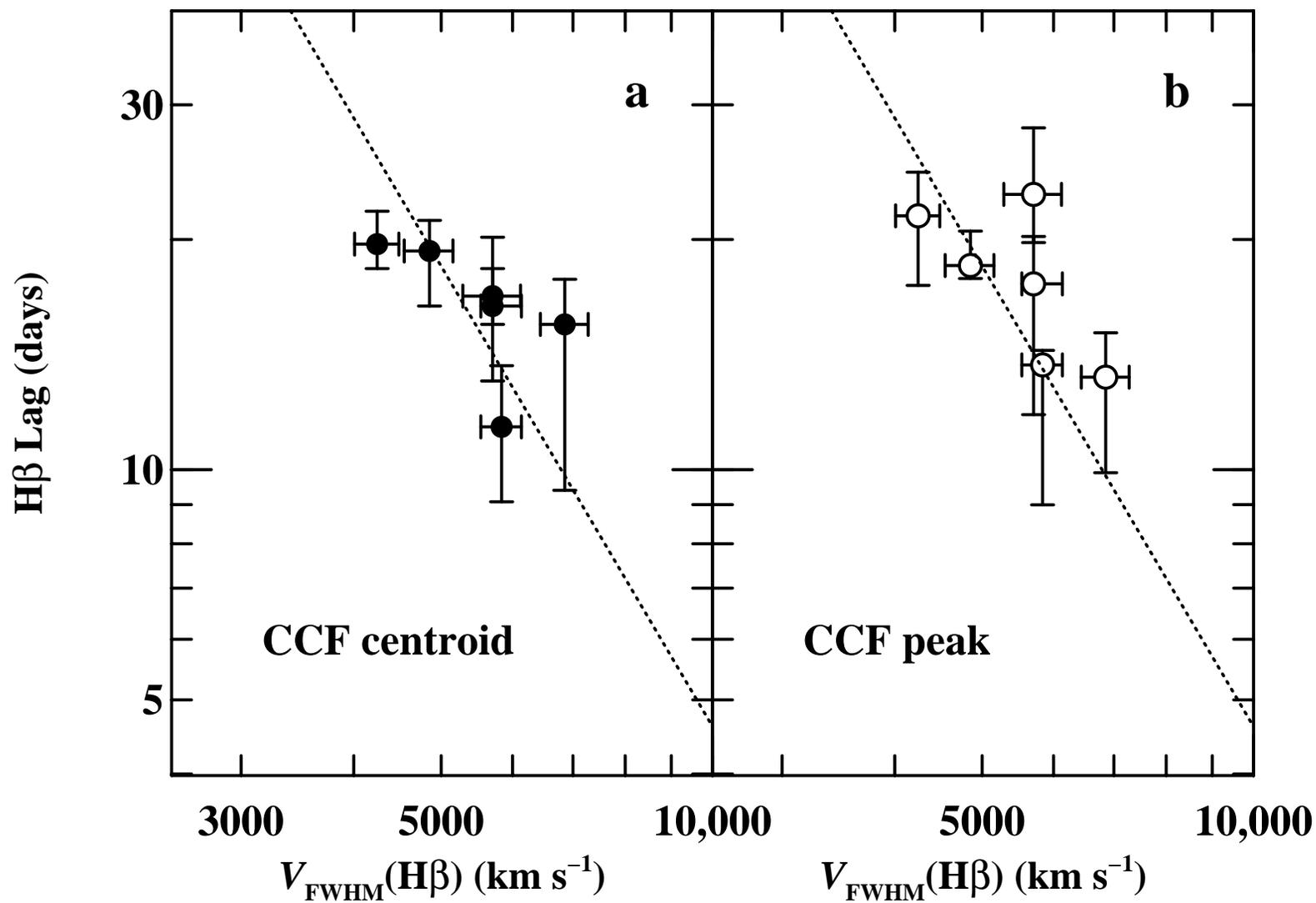
\end{document}